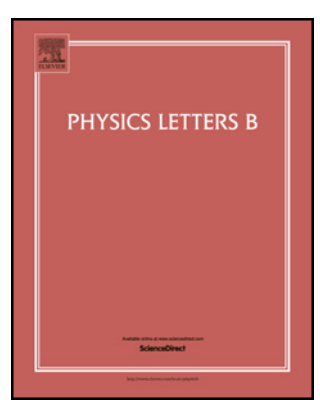

# Cosmic "adiabatic" photon creation: Temperature law and blackbody spectrum

J.A.S. Lima [a,*], S.R.G. Trevisani [a], R.C. Santos [b]

[a] *Universidade de São Paulo, Departamento de Astronomia (IAGUSP), Rua do Matão, 1226, 05508-900, São Paulo, SP, Brazil*
[b] *Universidade Federal de São Paulo (UNIFESP), Campus de Diadema, SP, Brazil*

ARTICLE INFO

ABSTRACT

The temperature-redshift relation (cosmic thermometer) is rediscussed in the framework of gravitationally induced photon production in the expanding universe. It is found that the same extended temperature law can be derived based on three independent methods, namely: (i) through a simple heuristic approach, (ii) based on macroscopic irreversible thermodynamics with photon creation, and (iii) using a kinetic relativistic Boltzmann equation incorporating gravitational photon creation. All these methods yield the same answer. The modified kinetic treatment reveals not only that the Planckian blackbody spectrum endowed with the "adiabatic" creation is preserved in the course of the expansion but also that CMB anisotropies and distortions can be treated properly. The new cosmic thermometer suggests a crucial test to the standard cosmic concordance model in the thermal sector. It is also argued that the present results may open a new route for scrutinising the $H_0$ Supernovae-CMB tension challenging the $\Lambda$CDM model.

The main pillars of the current standard cosmology ($\Lambda$CDM) are the prediction and successful observation of the cosmic microwave background (CMB) radiation, including its Planck spectral distribution, the abundance of the light elements and the discovery of the current accelerating stage of the Universe from Supernovae Ia (SN) observations. In addition, the perturbed $\Lambda$CDM solutions describe the structure formation process, its matter power spectrum and also the pattern of cosmic microwave background (CMB) temperature anisotropies (angular power spectrum). When combined with the early inflationary scenario, the resulting cosmic concordance model (inflation + $\Lambda$CDM) driven by the vacuum energy density ($\rho_v = \Lambda/8\pi G$), is considerably simple (flat geometry), attractive and predictive [1,2].

Nevertheless, beyond the theoretical difficulties related to the cosmological constant and coincidence problems [3,4] for which many solutions have been proposed [5–8], the $\Lambda$CDM model is now being observationally challenged by the current SN-CMB tension. In brief, local measurements of the Hubble constant ($H_0$) using Cepheid-calibrated supernovae of the SHOES collaboration [9], measurements at intermediate redshifts from combinations of tests [10], and strong lensing time-delays [11] provided a best fit value of $H_0 = 74$ km/s/Mpc while the Planck CMB data [2] points to a lower value close to 68 km/s/Mpc. The weak $2.6\sigma$ SN-CMB tension of a decade ago is now a clear-cut tension around ($5.4\sigma$) according to the latest results of SHOES and Planck collaborations.

In the theoretical front there is a tendency suggesting that the $H_0$ tension may potentially be relaxed due to systematic uncertainties related to dust extinction effects in the Cepheid colour-luminosity calibration associated to local measurements (see [12] and Refs. there in). It was also remarked that removing the current CMB temperature $T_0$ obtained by the FIRAS-COBE, a new window is open since the Planck data alone may be consistent with a broader region on the $T_0 - H_0$ plane [13]. Such very welcome attempts are trying to alleviate the SN-CMB tension (less discrepant $H_0$ values) with the extra bonus that all the successes of $\Lambda$CDM are preserved without introducing new physics.

Here we consider a less conservative viewpoint by discussing the loss of generality underlying the familiar temperature-redshift relation (TRR), sometimes dubbed Friedmann law, $T = T_0(1 + z)$. The value of $T_0$ has precisely been measured by the FIRAS-COBE satellite and recalibrated by the Wilkinson Microwave Anisotropy Probe (WMAP), $T_0 = 2.72548 \pm 0.00057$ K [14,15]. Interestingly, since the photon to baryon ratio is very high, the same temperature law remains valid at early times when the universe was opaque at high redshifts. Since long ago, the standard view is that such TRR is the only possibility preserving the blackbody form of the spectrum in the course of the cosmic expansion [16–19].

In a point of fact, there exist many reasons favouring the standard TRR. Macroscopically, it can be easily derived based on the conserved entropy of CMB in a comoving volume [$S \propto T^3 V = constant$], and also from a kinetic theoretic approach using the equilibrium phase space density, $f(t,p)$, or equivalently the photon "occupation" number, $n_\gamma(t,\nu) = (e^{h\nu/kT} - 1)^{-1}$. Further, the observed CMB spectrum is almost a perfect blackbody with a high degree of accuracy. This explains why the Friedmann law works like a theoretical thermometer determining the energy scale and the sequence of events in the cosmic history. It is also worth noticing that the standard TRR law and its associated Planckian spectrum are one of the basic ingredients to the kinetic perturbative treatment involving the generation of CMB anisotropies, spectral distortions, and their possible correlations [20].

* Corresponding author.
*E-mail addresses:* jas.lima@iag.usp.br (J.A.S. Lima), strevisanijr@usp.br (S.R.G. Trevisani), rose.santos@unifesp.br (R.C. Santos).

In this *Letter*, a more general cosmic thermometer is proposed in three steps. Firstly, the unicity of the standard relation is rediscussed through an heuristic argument suggesting an extended temperature law with photon creation. Secondly, by using an irreversible photon creation approach, we deduce the same modified temperature law heuristically suggested. Finally, a kinetic theoretic treatment based on the relativistic Boltzmann equation by taking into account "adiabatic" photon creation is proved to be exactly solved by the same temperature relation incorporating this kind of process.

The "adiabatic" terminology here means that the creation process is not isentropic. In this case, the comoving entropy increases, but the specific entropy (per photon) remains constant. Our key result at light of the Boltzmann equation with "adiabatic" photon creation is that the unperturbed distribution (zero order) is also described by a Planckian spectrum. In principle, such results may open a new window for new physics and also clarify how CMB anisotropies and other distortions would be rediscussed in this enlarged framework, potentially, bringing new light to the $H_0$ tension plaguing the $\Lambda$CDM model.

In what follows we consider that the spacetime geometry is described by a flat ($k=0$) Friedmann-Lemaître-Robertson-Walker (FLRW) metric:

$$ds^2 = dt^2 - a^2(t)\left[dx^2 + dy^2 + dz^2\right], \tag{1}$$

where $a(t)$ is the scale factor.

As a kind of warm-up, we first discuss the basic criticism for models with photon creation and its possible influence on the temperature law. Let us assume for a while that the standard equilibrium temperature relations for the energy and number densities, $\rho \propto T^4$, $n \propto T^3$, and $\rho \propto n^{4/3}$, are obeyed by an arbitrary homogeneous and isotropic *photon creation process*. Of course, such relations do not imply necessarily on the validity of a blackbody spectrum. As we shall see below, the nature of the specific spectrum satisfying these relations may be identified only by using more rigorous approaches.

The photon redshift relation due to the cosmic expansion is not modified. Actually, it is a pure kinematic geodetic effect ($ds^2 = 0$) over frequencies and wavelengths [16]

$$\frac{\lambda_0}{\lambda} = \frac{\nu}{\nu_0} = \frac{a_0}{a} = 1 + z. \tag{2}$$

Homogeneous and isotropic photon creation also implies that the average number of photons in comoving volumes, $N(t) = na^3$, are not conserved. Since we are assuming $n \propto T^3$, it follows that

$$\frac{\dot{T}}{T} + \frac{\dot{a}}{a} = \frac{1}{3}\frac{\dot{N}}{N} \;\leftrightarrow\; TaN(t)^{-1/3} = const. \tag{3}$$

Clearly, the standard law $aT = a_0 T_0$ is obeyed only if $N = const$. However, the nature of the thermodynamic process leading to the above expressions and also the form of the spectrum have not been determined. Hence, more general and rigorous macroscopic and kinetic derivations including photon creation are now required. In what follows we first extend the original formulation started by Prigogine et al. [21] and reanalysed in [22] through a manifestly covariant approach. Later on, a simple proof of the blackbody spectrum is given.

The nonequilibrium states of a radiation fluid ($p = \rho/3$) with gravitationally induced photon production and entropy are characterized by three independent macroscopic quantities: the energy-momentum tensor $T^{\mu\nu}$, particle current $N^\mu$, and entropy current $S^\mu$:

$$T^{\mu\nu} = (\rho + p)u^\mu u^\nu - (p + P_c)g^{\mu\nu}, \tag{4}$$

$$N^\mu = nu^\mu, \tag{5}$$

$$S^\mu = su^\mu, \tag{6}$$

where $\rho$, $p$ and $P_c$ are the energy density, thermostatic pressure and the dynamic (irreversible) creation pressure, respectively, while $n$ and $s$, are the particle number and entropy densities.

In the FLRW background, the evolution equations for such quantities take the form:

$$u_\mu T^{\mu\nu};_\nu \equiv \dot{\rho} + (\rho + p + P_c)\Theta = 0\,, \tag{7}$$

$$N^\mu;_\mu \equiv \dot{n} + n\Theta = n\Gamma_N, \;\leftrightarrow\; \frac{\dot{N}}{N} = \Gamma_N \tag{8}$$

$$S^\mu;_\mu \equiv \dot{s} + s\Theta = s\Gamma_S, \;\leftrightarrow\; \frac{\dot{S}}{S} = \Gamma_S \tag{9}$$

where the scalar of expansion $\Theta = 3H = 3\dot{a}/a$ ($H$ is the Hubble parameter) and (;) means covariant derivative. The quantities $\Gamma_N$ and $\Gamma_S$ denote the production rates of particles and entropy. Note the difference between the source terms on the balance equations, (8) and (9). In principle, there are two different rates and both are null for equilibrium states. The first one is clearly related to the emergence of particles in the spacetime, while the second one resembles the variation rate of the specific entropy, a typical contribution from bulk viscosity mechanism (second viscosity). In general $\Gamma_N \neq \Gamma_S$. However, its equality specifies a physically interesting case dubbed "adiabatic" creation [22] which is fully different from the bulk viscosity mechanism proposed long ago by Zeldovich [23] as an effective description of matter creation (more details in Lima and Germano [22]).

Let us recall that the fluid thermodynamic quantities are related to its temperature by the local Gibbs law:

$$nTd\left(\frac{s}{n}\right) \equiv nTd\sigma = d\rho - \left(\frac{\rho + p}{n}\right)dn, \tag{10}$$

where $\sigma = s/n = S/N$ is the specific entropy (per photon), while the chemical potential $\mu$ is defined by the so-called Euler relation, $n\mu = \rho + p - nT\sigma$. Note also that the above expression implies that the specific entropy in general is not conserved:

$$\dot{\sigma} = \sigma\left(\frac{\dot{S}}{S} - \frac{\dot{N}}{N}\right) \equiv \sigma\,(\Gamma_N - \Gamma_S)\,. \tag{11}$$

"Adiabatic" creation is characterised by $\dot{\sigma} = 0$, and occurs only if $\Gamma_N = \Gamma_S$. In this case, both the entropy and total number of particles increases but the specific entropy remains constant. In addition, one obtains the interesting result:

$$\frac{\dot{S}}{S} = \frac{\dot{N}}{N}, \quad \rightarrow S = k_B N + const., \tag{12}$$

where $k_B$ is the Boltzmann constant. The "adiabatic" entropy growth is actually closely related with the emergence of particles in the space-time thereby increasing at the same rate. An important point to keep in mind is that the presence of the creation pressure in this macroscopic description is not the result of collisional processes as happens, for instance, in the standard bulk viscosity mechanism. In what follows we consider only the "adiabatic" creation of photons which means that $\Gamma_S = \Gamma_N \equiv \Gamma$.

At this point, we need to determine the nonequilibrium photon creation pressure, $P_c$. First of all, it is easy to see that the "adiabatic" process satisfies the second law of thermodynamics. The Euler relation applied for photons ($\mu \equiv 0$), implies that $nT\sigma \equiv sT = \rho + p$. Now, since we are describing "adiabatic" photon creation, the balance equation for the entropy as given by (9) boils down to

$$S^{\mu};_{\mu} = n\sigma\Gamma = (\frac{\rho+p}{T})\Gamma \geq 0, \tag{13}$$

as required by the second law of thermodynamics. In addition, under "adiabatic" conditions, the Gibbs law (10) implies that

$$\frac{\dot{\rho}}{\rho+p} = \frac{\dot{n}}{n}, \tag{14}$$

and combining with the energy conservation law (7) and the balance equation for the number density (8), the photon creation pressure is obtained

$$P_c = -(\rho+p)\frac{\Gamma}{\Theta} = -\frac{4\rho}{3}\frac{\Gamma}{\Theta}. \tag{15}$$

This simple expression remains valid as long as the "adiabatic" condition is maintained. Note also that an integration of (14) yields the equilibrium relation, $\rho \propto n^{4/3}$ thereby suggesting the following questions: *Are the remaining equilibrium relations ($n \propto T^3$, $\rho \propto T^4$) also obeyed as assumed in the heuristic derivation? How the radiation temperature evolution law is modified in this context?*

In order to answer such questions let us assume $\rho = \rho(T, n)$, $p = p(T, n)$ and the well known thermodynamic identity:

$$T\left(\frac{\partial p}{\partial T}\right)_n = \rho + p - n\left(\frac{\partial \rho}{\partial n}\right)_T. \tag{16}$$

Now, with the creation pressure given by (15) one may rewrite the energy conservation law (7) in terms of $T$, $n$ and $\Gamma$. Furthermore, by using the condition (14), the balance equation for the number density (8) and the above identity, the temperature evolution of radiation with creation reads:

$$\frac{\dot{T}}{T} = \left(\frac{\partial p}{\partial \rho}\right)_n \frac{\dot{n}}{n} = \frac{1}{3}\frac{\dot{n}}{n} \;\Leftrightarrow\; \frac{\dot{\rho}}{\rho} = 4\frac{\dot{T}}{T}. \tag{17}$$

Hence, as assumed without proof in the heuristic derivation, we see that $n \propto T^3$, $\rho \propto T^4$ and $\rho \propto n^{4/3}$. All the temperature equilibrium relations for $\rho$, $n$ and $\sigma$ are effectively satisfied in the case of "adiabatic" photon creation. Interestingly, the heuristic derivation present in the beginning of this section is also encoded in the above temperature law. By combining (17) with (8) under "adiabatic" condition, we find

$$\frac{\dot{T}}{T} = -\frac{\dot{a}}{a} + \frac{\Gamma}{3} = -\frac{\dot{a}}{a} + \frac{1}{3}\frac{\dot{N}}{N} \;\leftrightarrow\; TaN^{-1/3} = const., \tag{18}$$

where in the second equality $\Gamma = \dot{N}/N$ was replaced. This is exactly the same temperature law derived from heuristic arguments presented in the warm-up of this macroscopic irreversible approach [compare with (3)].

It should be also stressed the generality of such temperature law. In particular, by using the above equilibrium relations, it is readily checked that all balance equations yield the same temperature law. It can also be deduced from the entropy function $S \propto T^3V$ combined with the entropy balance equation, $\dot{S} = \Gamma S$. Therefore, the temperature law with creation is not only more general than the Friedmann law, $Ta = constant$, but also retains the same degree of universality.

It is also widely believed that the cosmological creation of photons in the expanding Universe is not allowed because the blackbody form of the CMB spectrum should be inevitably destroyed. Hence, it is natural to ask: *what is the nature of the spectrum for "adiabatic" photon creation?* Actually, such a question was first discussed long ago, but it was not completely settled because the lack of a properly modified Boltzmann equation by taking into account photon creation [24]. In what follows, a comprehensive and more complete answer including how to treat distortions in this context is provided based on a properly modified Boltzmann equation.

Let us now focus our attention on the kinetic counterpart of the irreversible macroscopic approach for gravitational "adiabatic" photon production as above discussed. The main goal now is to show from first principles that under "adiabatic" conditions the Planckian spectrum is also preserved in the course of the expansion. As should be expected, this happens exactly due to the validity of (18), the extended temperature law.

The radiation field is described by the photon phase space density $f(x^\mu, p^\mu)$, or equivalently, by $n_\gamma(t, \nu)$. The evolution of both forms is governed by the same relativistic kinetic Boltzmann equation but now including photon creation. Such an equation was proposed a couple of years ago [25].

$$p^\mu \frac{\partial f}{\partial x^\mu} - \Gamma^\mu_{\alpha\beta} p^\alpha p^\beta \frac{\partial f}{\partial p^\mu} = C(f) + \bar{C_p}(f), \tag{19}$$

where $\Gamma^\mu_{\alpha\beta}$ are the Christoffel symbols and C(f) encompass the collisional contributions describing elastic and inelastic processes. The extra $\bar{C_p}(f)$ quantity is a phenomenological non-collisional source term encoding the "adiabatic" photon production process due to the cosmic time-varying geometry. Such a term of gravitational origin is quite distinct from any collisional contribution present in the standard treatment [25]. It can also be interpreted as a new contribution of gravitational origin in the left hand side of the Boltzmann equation. More precisely, an irreversible energy flow from the gravitational field to the created matter constituents, a viewpoint closely related to quantum field theory in curved spacetimes [26] and thermodynamics of open systems in cosmology [21,22].

By neglecting standard collisions [$C(f) \equiv 0$], the extended Liouville operator on the mass shell now reads:

$$L(f) = \frac{\partial f}{\partial t} - \frac{\dot{a}}{a} p \frac{\partial f}{\partial p} + \frac{\Gamma}{3} p \frac{\partial f}{\partial p} = 0. \tag{20}$$

Note that for $\Gamma << H$, the ratio $\Gamma/3H$ is negligible and the standard Liouville operator is recovered.

In terms of the occupation number, $n_\gamma(t, \nu)$, the above equation can also be rewritten as

$$\frac{\partial n_\gamma}{\partial t}|_\nu - \nu H \frac{\partial n_\gamma}{\partial \nu}|_t + \frac{\Gamma}{3} \nu \frac{\partial n_\gamma}{\partial \nu}|_t = 0. \tag{21}$$

For the sake of generality, let us now consider a radiation gas of massless bosons or fermions with $\mu = 0$ (henceforth we consider units with $h = k_B = c = 1$). The equilibrium phase space densities for massless quanta can be written as [27]

$$n_\gamma(t, \nu) = \frac{1}{exp(\frac{\nu}{T}) - \epsilon}, \tag{22}$$

where ($\epsilon = +1$) for bosons and ($\epsilon = -1$) for fermions.

In order to obtain the condition under which this equilibrium phase space density is a solution of the kinetic Boltzmann type equation (21), it is convenient to define the quantity $B^2 = [exp(\frac{\nu}{T}) - \epsilon]^{-2}$. As one may check

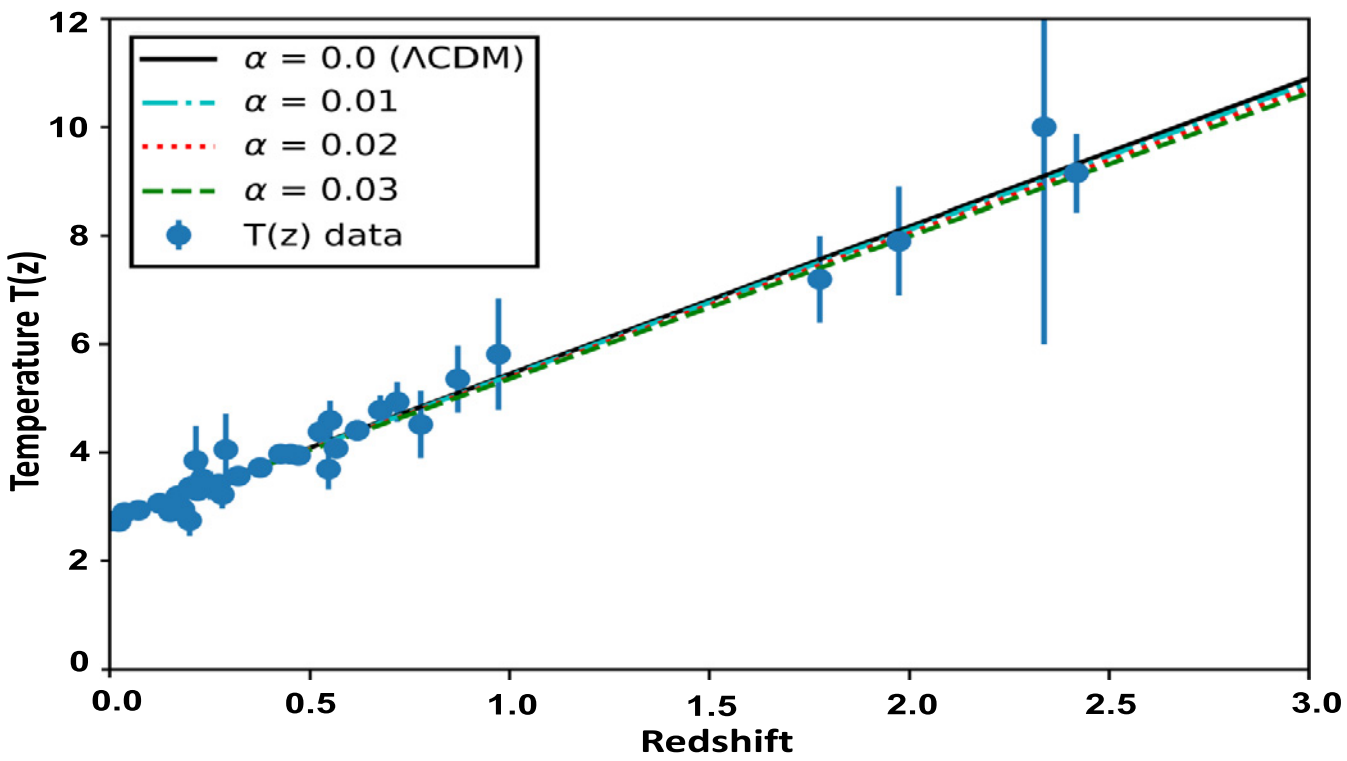

**Fig. 1.** $T(z)$ for constant photon creation rate. The solid black line ($\alpha = 0$) corresponds to the standard linear temperature law ($\Gamma_0 = 0$). For $\alpha$ and z different from zero, the law is non-linear an has a dependence on the Hubble constant $H_0$ [see (27)]. Such a connection is absent in the linear law and may help to alleviate the current tension on $H_0$.

$$\frac{\partial n_\gamma}{\partial t}|_\nu = B^2 \frac{\nu}{T} e^{\frac{\dot{T}}{T}}, \tag{23}$$

$$\nu \frac{\partial n_\gamma}{\partial \nu}|_t = -B^2 \frac{\nu}{T} e^{\nu/T}|. \tag{24}$$

Now, inserting these results into (21) we obtain:

$$B^2 \frac{\nu}{T} e^{\nu/T} \left[ \frac{\dot{T}}{T} + \frac{\dot{a}}{a} - \frac{\Gamma}{3} \right] = 0 \leftrightarrow \frac{\dot{T}}{T} = -\frac{\dot{a}}{a} + \frac{\Gamma}{3}, \tag{25}$$

a result in perfect agreement with the heuristic and macroscopic deductions [see (3) and (18)]. The required condition for a preserved Planck spectrum in the course of the expansion is just the validity of the extended temperature law with "adiabatic" photon creation.

It should be stressed that distortions in the Planckian spectrum with "adiabatic" creation can readily be investigated by adding the conventional source terms to the extended collisionless Boltzmann equation (21). In this case, the equation governing the evolution of the occupation number takes the form:

$$\frac{\partial n_\gamma}{\partial t}|_\nu - \nu H \frac{\partial n_\gamma}{\partial \nu}|_t + \frac{\Gamma}{3} \nu \frac{\partial n_\gamma}{\partial \nu}|_t = \frac{dn_\gamma}{dt}|_C + \frac{dn_\gamma}{dt}|_{DC} + \frac{dn_\gamma}{dt}|_{BR}, \tag{26}$$

where the terms on the r.h.s. are the most common sources of distortions, namely, Compton scattering, double Compton and Bremsstralung emission, respectively (the above equation should be compared with Eq. (1) of [28] where a different approach to treat distortions in the present context was discussed). Note that the spectrum is Planckian but only when these sources are not taken into account. Therefore, it is clear that gravitationally induced "adiabatic" photon creation by itself does not provoke any distortion.

Distortions and its connections with the CMB anisotropies is a current hot topic in the literature [20,29,30]. Several projects have been proposed for observing CMB spectral distortions at the sensitivity required to probe this well known phenomenology (see, for instance, Astro2020 APC White Paper [31]). In this context, "adiabatic" photon creation as discussed here is a clear possibility of new physics. Investigations along these lines are in progress, however, a more detailed treatment is out of scope of this *Letter*.

In order to illustrate the photon creation effect on the temperature law, let us assume a constant creation rate since recombination ($\Gamma \equiv \Gamma_0$) and a $\Lambda$CDM evolution. In this case the comoving number of photons and temperature read:

$$N(z) = N_0 e^{-\Gamma_0(t_0 - t(z))} \rightarrow T = T_0(1+z) e^{-\alpha F(\Omega_m, z)}, \tag{27}$$

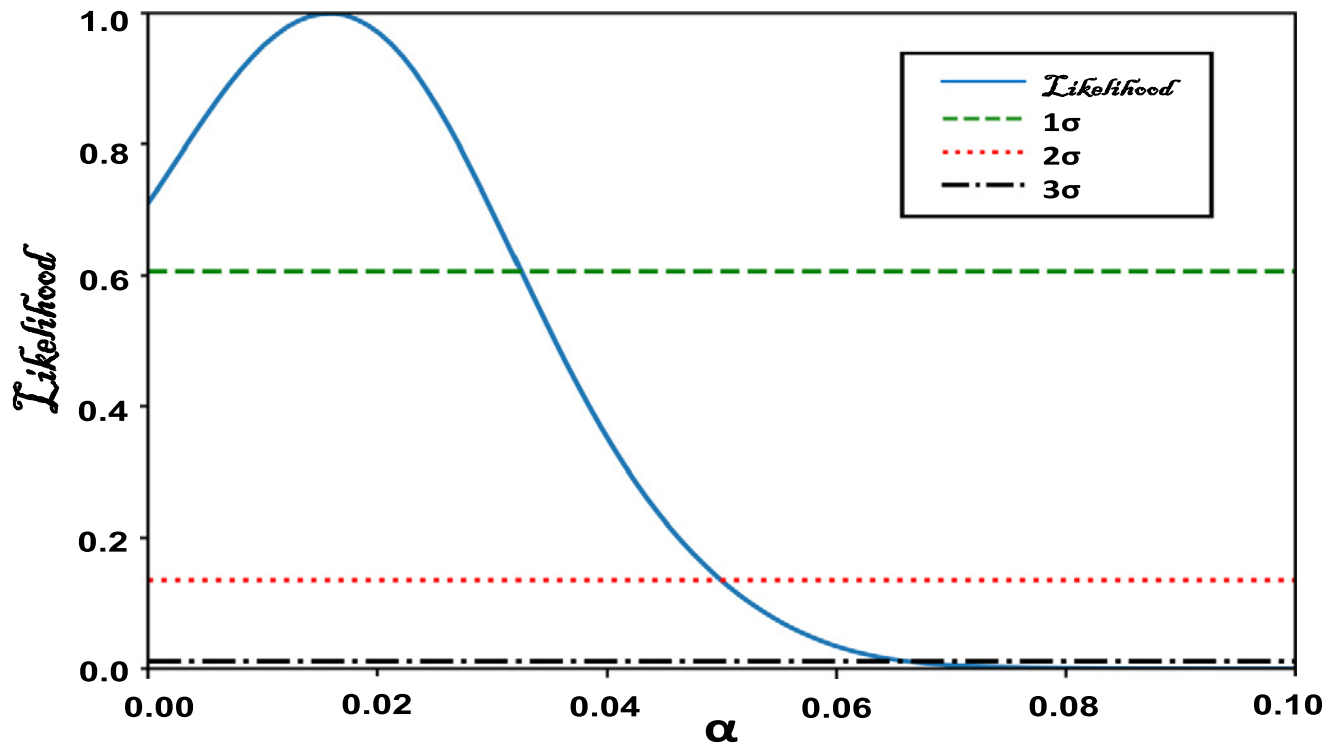

**Fig. 2.** Likelihood for constant photon creation rate. The second law of thermodynamics constrains $\Gamma_0$ to positive values. This means that the free parameter $\alpha = \Gamma_0/3H_0$ is also positive. Using the best fit of $\alpha$ (see text), we find $\Gamma_0 \simeq 0.04 H_0$.

where $N_0$, $T_0$ are present day values, $\alpha = \Gamma_0/3H_0$ and $F(\Omega_m, z) = \int_{\frac{1}{1+z}}^{1} [\Omega_m x^{-1} + (1-\Omega_m)x^2)]^{-1/2} dx$. As expected, the photon creation process implies that $N(z) \le N_0$. Both quantities $N(z)$ and $T(z)$ depend on the cosmic lookback time $[t_0 - t(z)]$.

Fig. 1 displays the function $T(z)$ for some selected values of the free parameter $\alpha$. Measurements of the CMB temperature are usually obtained from Sunyaev-Zeldovich [32] effect and also through excitation states of interstellar molecules (e.g., C, CO, CNO). All T(z) data were compiled from Refs. [33–40]. For a given value of $z \neq 0$, the temperature is smaller than in the standard $\Lambda$CDM model with no creation (solid black line). For $\alpha$ and $z$ different from zero, the law is non-linear and has a dependence on the Hubble constant $H_0$ [see (26)]. Such a connection is absent in the linear TRR and may help to alleviate the current tension on the values of $H_0$.

In Fig. 2, for the same set of data, we also show the likelihood of the dimensionless $\alpha$-parameter defined in terms of the creation rate $\Gamma_0$. In our analysis $\alpha = 0.0114^{+0.0174}_{-0.0176}$ at $1\sigma$ confidence level.

It is also natural to ask about the underlying physical process involving "adiabatic" gravitational creation of massless particles as discussed here. Particle production in an expanding Universe was investigated by Parker and others using the Bogoliubov mode-mixing technique in the context of quantum field theory (QFT) in curved spacetime (see [26] and Refs. therein). A classical field when quantized in an expanding background (e.g., the flat geometry of $\Lambda$CDM) implies the existence of particle creation with the energy for the newly created particles supplied by the classical time-varying geometry. Hence, assuming that the macroscopic and kinetic results derived here will be confirmed by independent calculations, the implicit connection with the methods from QFT in curved spacetimes becomes an interesting and challenging subject.

Summing up, we have investigated the possibility of an extended cosmic thermometer (CMB temperature law) by taking into account the concept of "adiabatic" photon creation. The CMB temperature law describing the cosmic history at the unperturbed level was derived based on three different paths, namely: heuristic, macroscopic and a kinetic theoretic approach.

The "adiabatic" concept here means that entropy increases in the course of time, but the specific entropy remains constant during the photon creation process. This happens when the creation rates of particles and entropy are the same, $\dot{S}/S = \dot{N}/N = \Gamma$, or equivalently, when the creation rates of particles and entropy are equal [see discussions below Eq. (11)].

The spectrum in the "adiabatic" creation process for massless particles is proved to be described by the equilibrium Planckian and Fermi-Dirac distributions for bosons ($\epsilon = 1$) and fermions ($\epsilon = -1$), respectively. When the creation rate is negligible (e.g., $\Gamma_0 \sim 10^{-5} H_0$), the only difference is that the standard temperature law describing the true adiabatic process in the expanding uni-verse is recovered. All independent deductions of the temperature law with "adiabatic" creation as presented here suggest that standard calculations describing distortions or CMB anisotropies must be carried out by adding the well known collisional operators over the gravitationally modified Liouville operator [see Eqs. (21) and (26)]. The above considerations are valid not only for the gravitationally induced photon creation, but also when two components are interchanging energy and particles obeying the "adiabatic" condition [33,41].

The general case ($\Gamma_S \neq \Gamma_N$) does not satisfy the "adiabatic" condition because $\dot{\sigma} \neq 0$. Entropy production in this case is also accompanied by the production of specific entropy (per photon). However, the calculated creation pressure and the temperature law are modified. Preliminary results suggest that the blackbody spectrum is not preserved in the course of the expansion [42].

Finally, it is worth noticing that the "adiabatic" photon creation process and the associated temperature law imply that the pattern of CMB anisotropies will be slightly modified. This means that even the integrated Sachs-Wolfe effect as predicted by the cosmic concordance ($\Lambda$CDM) cosmology must change in comparison with the standard approach (no 'adiabatic' creation). Potentially, the present approach opens a new window for scrutinizing not only the SN-CMB $H_0$ tension but also other tenets of the current $\Lambda$CDM cosmology.

## Declaration of competing interest

The authors declare that they have no known competing financial interests or personal relationships that could have appeared to influence the work reported in this paper.

## Acknowledgements

The authors are grateful to J. F. Jesus for helpful discussions. JASL is partially supported by the National Council for Scientific and Technological Development (CNPq) under grant 310038/2019-7, CAPES (88881.068485/2014), and FAPESP (LLAMA Project No. 11/51676-9). SRGT also acknowledges a fellowship from CNPq under grant 142502/2018-9.